\def\USEACHEMSO{0} %

\if\USEACHEMSO1
\documentclass[journal=jpclcd,manuscript=perspective,%
email=true,%
]{achemso}
\else
\documentclass[aip,notitlepage,%
jcp,
tightenlines,
twocolumn,
reprint,%
floatfix]{revtex4-2} 
\fi
\usepackage{orcidlink}

\usepackage[utf8]{inputenc}
\usepackage[T1]{fontenc} 
\usepackage{stix}
\usepackage[version=3]{mhchem}
\usepackage{hyperref}
\usepackage{units}

\usepackage{booktabs}
\usepackage{dcolumn}%

\usepackage{xspace}
\usepackage{verbatim}
\let\oldtheequation\theequation
\makeatletter
\def\tagform@#1{\maketag@@@{\ignorespaces#1\unskip\@@italiccorr}}
\renewcommand{\theequation}{(\oldtheequation)}
\makeatother 

\graphicspath{{pics}}

\usepackage{listings}

\usepackage[textsize=small,textwidth=1.2cm]{todonotes}

\newcommand{\matr}[1]{\ensuremath{\mathbf{#1}}}
\newcommand{\braket}[2]{\ensuremath{ \langle #1 | \, #2  \rangle }}

\newcommand{\ketbra}[2]{\ensuremath{  | {#1} \rangle \langle {#2} |}}

\newcommand{\ket}[1]{\ensuremath{  | {#1} \rangle}}

\newcommand{\matrixe}[3]{\ensuremath{ \langle{#1} | {#2} | {#3} \rangle }}

\newcommand{\dd}{\ensuremath{\mathrm{d}}}
\newcommand{\ii}{\ensuremath{\mathrm{i}}}

\newcommand{\lit}[1]{Ref.~\mbox{[\!\!\citenum{#1}]}\xspace}

\newcommand{\icm}{cm^{-1}}

\newcommand{\bdim}{\ensuremath{D}}
\newcommand{\bdimmax}{\ensuremath{D_\text{max}}}

\definecolor{CBdblue}{RGB}{5,113,176}

\if\USEACHEMSO0
\begin{document}
\fi

\title{Accurate, full-dimensional computations of thousands of complex vibrational eigenstates with tree tensor network states}%

\author{Henrik R.~Larsson}%
\affiliation{Department of Chemistry and Biochemistry, University of California, Merced, CA 95343, USA}
\if\USEACHEMSO0
\affiliation{Department of Physics, University of California, Merced, CA 95343, USA}
\else
\altaffiliation{Department of Physics, University of California, Merced, CA 95343, USA}
\fi
\if\USEACHEMSO1
\else
\email{pTTNS426 [$\alpha \tau$] larsson-research . $\delta\epsilon$}
\fi
\author{Brieuc Le D\'e~}%
\affiliation{Department of Chemistry and Biochemistry, University of California, Merced, CA 95343, USA}
\author{Gino E. Gamboni}%
\if\USEACHEMSO0
\affiliation{Department of Chemistry and Biochemistry, University of California, Merced, CA 95343, USA}
\else
\altaffiliation{Department of Chemistry and Biochemistry, University of California, Merced, CA 95343, USA}
\fi
\affiliation{Department of Physics, University of California, Merced, CA 95343, USA}

\begin{abstract}
Tree tensor network states (TTNSs) combined with the density matrix renormalization group (DMRG) are emerging as powerful tools for vibrational and vibronic structure simulations in molecules with strong coupling and fluxionality.
In this Perspective, we discuss how TTNS methods enable accurate, full-dimensional computations of thousands of eigenstates for molecular systems ranging  from quartic-force-field benchmarks to molecules with strong vibronic coupling and protonated water clusters as large as the 33-dimensional Eigen ion, \ce{H3O+.(H2O)3}. 
We emphasize the close connection and interoperability between DMRG-based TTNS methods and the multilayer multiconfiguration time-dependent Hartree method (ML-MCTDH), which share the same underlying ansatz. 
We also highlight practical challenges of 
predictive simulations, 
including robust error estimation, convergence of observables such as infrared intensities, and optimization of tensor network tree structures. 
Finally, we outline recent advances toward direct targeting of excited states and discuss opportunities for broader applications in molecular spectroscopy and quantum dynamics.
\end{abstract}

\maketitle

Nuclear quantum effects are important for understanding and predicting the properties of molecules and matter.\cite{Dynamics2007vendrell,Strong2009vendrell,Imaging2015vogels,Quantuminduced2010ivanov,Nuclear2016ceriotti,Hydrogen2014layfield,Quantum2021chen,Quantum2025nguyen}
Vibrational spectra help in decyphering these effects and characterize the quantum dynamics of atoms in molecules.\cite{Dynamics2007vendrell,Strong2009vendrell,HighLevel2017yu,Disentangling2017duong,Breaking2025jing,Numerically2025wanga,H2O2025simko}
One important example where nuclear quantum effects are crucial for understanding the molecular dynamics and vibrational spectra are protonated water clusters, which 
help us in characterizing properties of the hydrated proton and 
reveal chemical phenomena such as bonding, (micro-)solvation and acidity,
as well as quantum phenomena such as resonances, tunneling and zero-point energy.\cite{Spectral2005headrick,Proton2006marx,Dynamics2007vendrell,Strong2009vendrell,Snapshots2015fournier,Demystifying2021zeng,Crossover2021dereka}
The Zundel, \ce{H+.(H2O)2}, and the Eigen, \ce{H3O+.(H2O)3}, ions are particularly interesting, as they are regarded as possible building blocks of acidic water.\cite{Quantum1997tuckerman,Nature1999marx,Sequential2005mohammed,Spectroscopic2016wolke,Largeamplitude2017dahms,Resolving2021calio} 

Many methods exist to simulate vibrational spectra. Among others, these include 
vibrational perturbation theory,\cite{Vibrational2007christiansen,How2021frankea}
configuration interaction (VCI),\cite{AVCI2017odunlami,Vibrational2023trana,Automated2024schroder,Computational2026qu} %
coupled cluster,\cite{Vibrational2004christiansen,Similaritytransformed2018faucheaux,Vibrational2020klinting}
direct-product bases,\cite{Fourth2012csaszar,Calculated2016wang}
non-direct-product bases,\cite{Highly1986bacic,Computing2009cooper,Accurate2012halverson,Efficient2016larsson,Resonance2018larsson}
and 
sparse cubature.\cite{Calculating2006degani,Nonproduct2009avila,Quantum2014lauvergnat,Variational2024sunaga}  %
The fluxionality of the protonated water clusters and similar molecules leads to a multiconfigurational character of the wavefunctions, making them difficult to simulate. Consequently, methods that describe this character are well-suited for studying fluxional molecules.
Next to some of the aforementioned methods, 
tensor network methods such as 
those based on hierarchical canonical decompositions,\cite{Calculating2014leclerc,Using2018thomas}
the 
multilayer
multiconfiguration time-dependent Hartree method (ML-MCTDH),\cite{Multiconfigurational1990meyer,Multilayer2003wang,Multilayer2008manthe,Multilayer2011vendrell,Multilayer2015wang,Wavepacket2017manthe,Dynamical2017larsson,Molecular2026gatti}%
and the density matrix renormalization group (DMRG)\cite{Density1992white,Densitymatrix1993white,Vibrational2017baiardi,Computing2019larsson,Tensor2021glaser,Tensor2024larsson}
are particularly well-suited for simulating large, complex fluxional molecules in full dimensionality.
While established independently, ML-MCTDH uses the same ansatz as DMRG-based tree tensor network state (TTNS) methods, which stem from condensed-matter physics and electronic structure.\cite{Classical2006shi,Low2012chan}

Recently, we have shown that DMRG-based TTNS approaches
enable the computation of thousands of 
vibrational and vibronic eigenstates to very high accuracy.\cite{Computing2019larsson,Stateresolved2022larsson,vibronic2024larsson,Benchmarking2025larsson,Computing2025rano}
Eigenstates provide direct insights into the infrared (IR) spectrum that time-propagation-based methods such as MCTDH cannot. Using the time-dependent DMRG,\cite{Time2015lubicha,Unifying2016haegeman}
which is 
a particular way to solve the ML-MCTDH equations of motion, also known as the projector splitting integrator,\cite{Projectorsplitting2014lubich,Tensor2019schroder,Time2021lindoy} %
TTNSs can also be used for dynamics, which is useful for getting overview spectra.  %
Since TTNSs share the exact same ansatz used in ML-MCTDH, DMRG-based TTNS theory is fully compatible with ML-MCTDH theory and hence provides a complementary approach and full interoperability  with ML-MCTDH methods.

In this Perspective, we review our DMRG-based TTNS approach
and showcase how it can be used to compute thousands of highly accurate, full-dimensional eigenstates of real-world, non-model systems. 
Our examples range from molecules described by simple quartic force fields to very nonadiabatic vibronic systems and, particularly,
protonated water clusters as large as the 33-dimensional (33D) Eigen ion.
A central motive of this Perspective is highlighting difficulties in accurate eigenstate simulations of these molecular systems, such as ensuring reliable error estimates and converged observables beyond energies, in this case  IR intensities.
As such, this Perspective complements our recent review of ML-MCTDH theory through the lens of tensor network state theory.\cite{Tensor2024larsson}

\if\USEACHEMSO1
\paragraph*{Recap of Tree Tensor Network State Methods}
\else
\section*{Recap of Tree Tensor Network State Methods}
\fi

\begin{figure*}[!htbp]
  \includegraphics[width=.8\textwidth]{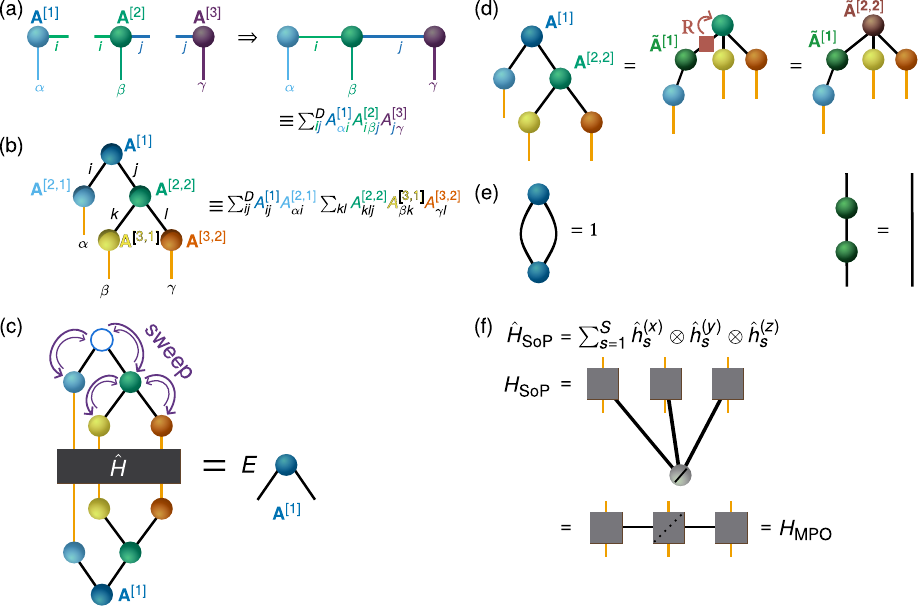}
    \caption{Tree tensor network state (TTNS) methodology.  (a) Example of a tensor network diagram     for a matrix product state (MPS)
    and the corresponding summation pattern. Compare with \autoref{eq:mps}.
    Nodes correspond to tensors and vertices/bonds denote their dimensions; shared vertices indicate summation  over the corresponding indices ($i$, $j$), while the free-standing vertices means that we do not sum over the corresponding indices ($\alpha$, $\beta$, $\gamma$).
    (b) Same as (a) but for a TTNS.
    (c) Eigenvector problem to compute an update for the root note $\matr A^{[1]}$ in (b).
  The large black tensor represents the Hamiltonian,
  whose typical tensor network structure is shown in (f).
A vectorized $\matr A^{[1]}$ corresponds to the eigenvector while the rest of the diagram on the left-hand-side corresponds to an effective Hamiltonian matrix.
The updated $\matr A^{[1]}$ replaces its predecessor in the TTNS (empty circle). 
The eigenvalue problem is repeated for all tensors in one ``sweep,'' which is indicated by the purple arrows. %
    (d) Canonicalization procedure that changes
    the orthogonality conditions for $\matr A^{[1]}$ and $\matr A^{[2,2]}$, leading to a new root node.
    The first step corresponds to reshaping $\matr A^{[1]}$ as matrix and QR-decomposing it. %
    The reshaped $\matr Q$ matrix is the new tensor  $\matr{\tilde A^{[1]}}$.
    The second step corresponds to the absorption of the $\matr R$ matrix from the QR decomposition into $\matr A^{[2,2]}$, leading to the tree shown on the right-hand side.
(e)  Orthogonality conditions of $\matr A^{[1]}$ 
before and after the canonicalization shown in (d).
The straight line corresponds to a unit matrix.
(f) Sum-of-products (SoP) Hamiltonian as a diagram. 
The one-dimensional Hamiltonians are shown as rectangular nodes. The circle with a diagonal corresponds to a unit tensor, i.e., $\delta_{s \tilde s}\delta_{s \overline s}$, turning $\sum_{s \tilde s \overline s}  [h_{s}^{(x)}]_{\alpha \alpha'} [h_{\tilde s}^{(y)}]_{\beta \beta'} [h_{\overline s}^{(z)}]_{\gamma \gamma'}$ into $\sum_{s} [h_s^{(x)}]_{\alpha \alpha'} [h_{s}^{(y)}]_{\beta \beta'} [h_{s}^{(z)}]_{\gamma \gamma'}$.
The SoP Hamiltonian corresponds to a matrix-product operator (MPO) with diagonal tensors (dotted, diagonal line). 
}
  \label{fig:tensor_overview}
\end{figure*}

To introduce TTNSs, 
we use a three-dimensional example, which can easily be generalized to higher dimensions.
In this example, the 
eigenstate $\ket{\Psi}$ is expressed by a direct-product basis of one-dimensional basis states $\ket{x_\alpha}$, $\ket{y_\alpha}$, and $\ket{z_\alpha}$ with basis sizes $N_x$, $N_y$, and $N_z$, respectively.
Commonly, these bases are given by grid-based discrete variable representations (DVR).\cite{Discretevariable2000light,PhaseSpace2018tannor}
The state is then given as
\begin{equation}
  \ket{\Psi} = \sum_{\alpha=1}^{N_x}  \sum_{\beta=1}^{N_y}  \sum_{\gamma=1}^{N_z}
  C_{\alpha \beta\gamma} \ket{x_\alpha y_\beta z_\gamma},
  \label{eq:fci}
\end{equation}
where all of the information about the state is encoded in the real-valued coefficient tensor $\matr C$.
While this ansatz works well for three-dimensional systems, the exponential scaling of the size of $\matr C$ renders direct applications to high-dimensional systems impossible.
An approximation that, for some systems,\cite{Multilayer2008manthe,Low2012chan,Colloquium2010eisert,Tree2025barthel}
alleviates the exponential scaling is to express $C_{\alpha \beta\gamma}$ in terms of summations over smaller-dimensional tensors. For example, in a matrix product state (MPS) or tensor train, we use the approximation
\begin{align}
  C_{\alpha \beta\gamma} &\approx 
   \sum_{ij} A^{[1]}_{\alpha i} %
  A^{[2]}_{i \beta j}  A^{[3]}_{j \gamma},\label{eq:mps}
\end{align}
where, for fixed indices $\alpha$, $\beta$, $\gamma$, the remaining entries of the tensors  $\matr A^{[\ell]}$ can be viewed as vectors (for the first and last dimensions) and matrices (for all dimensions but the first and last one, in this case just $\matr A^{[2]}$), %
hence the name MPS. 
Consequently, a four-dimensional MPS leads to
  $C_{\alpha \beta\gamma\delta} \approx 
   \sum_{ijk} A^{[1]}_{\alpha i} %
  A^{[2]}_{i \beta j}  A^{[3]}_{j \gamma k} A^{[4]}_{k \delta}$.
The sizes of these matrices are called bond dimensions, $\bdim_i$, and the maximum bond dimension is $\bdimmax$.
ML-MCTDH practitioners call $\bdim_i$ the number of single-particle functions, $n_\text{SPF}$.
The larger $\bdim_i$, the better the approximation.
The matrices in an MPS connect each dimension with their nearest neighbors and thus encode entanglement and correlation between each dimension.
Such an ansatz works particularly well for quasi-one-dimensional systems.
For vibrating molecules, however, the state is better described by an ansatz that allows for more coupling between not just nearest neighbors but groups of correlated vibrational modes.
Such an ansatz is given by a TTNS, which corresponds to nested or hierarchical summations of tensors, e.g., 
\begin{align}
  C_{\alpha \beta\gamma} &\approx 
   \sum_{ij} A^{[1]}_{ij} %
  A^{[2,1]}_{\alpha i} \sum_{kl} A^{[2,2]}_{klj} A^{[3,1]}_{\beta k} A^{[3,2]}_{\gamma l}.\label{eq:ttns}
\end{align}
Note that  $A^{[1]}_{ij}$ and $A^{[2,2]}_{klj}$ do not carry physical indices and rather mediate correlations between up to three degrees of freedom, instead of two for an MPS.
\autoref{eq:ttns} is a complex expression even for this three-dimensional example, to which we introduce a diagrammatic notation shown in \autoref{fig:tensor_overview}, with panel (a) showing an MPS and panel (b) showing a TTNS. Therein, each tensor corresponds to a node, and each index corresponds to a bond. 
Following Einstein's summation notation,\cite{Grundlage1916einstein} 
we sum over bonds/indices that are shared between two nodes/tensors. 
The diagram then leads to tree-like structures, thus the name TTNS.  The labels  in $\matr A^{[\ell,h]}$ correspond to the layer $\ell$ and the horizontal position $h$ in that layer in the tree.
The tensor (node) at the first layer, $\ell=1$ is dubbed root tensor (root node).
Note that these diagrams can be used to derive equations, which we used  in \lit{Tensor2024larsson} to derive the ML-MCTDH equations of motions.

To compute ground states, 
we 
plug our TTNS approximation of $C_{\alpha\beta\gamma}$,
\autoref{eq:ttns},
into the direct-product expansion, \autoref{eq:fci}, which yields 
\begin{equation}
  \ket{\Psi} = \sum_{ij}  A^{[1]}_{ij} \ket{\Phi_{ij}},
  \label{eq:ttns_rootexpansion}
\end{equation}
where we hide all but the root tensor $A^{[1]}_{ij}$ in the configurations that are defined as
\begin{equation}
  \ket{\Phi_{ij}} =
  A^{[2,1]}_{\alpha i}  \sum_{kl} A^{[2,2]}_{klj}  A^{[3,1]}_{\beta k} A^{[3,2]}_{\gamma l} 
  \ket{x_\alpha y_\beta z_\gamma}.
\label{eq:configs}
\end{equation}
$\ket{\Phi_{ij}}$
can be made orthogonal by enforcing that 
\begin{equation}
\sum_\alpha A^{[2,1]}_{\alpha i} A^{[2,1]}_{\alpha \tilde i} = \delta_{i\tilde i}, \quad \sum_{kl} A^{[2,2]}_{klj} A^{[2,2]}_{kl\tilde j} =\delta_{j\tilde j}, \text{etc.} %
\label{eq:orth_cond}
\end{equation}
These conditions are shown implicitly by the tree diagram in \autoref{fig:tensor_overview}(b), as those indices that lead to $\delta_{i\tilde i}$ in \autoref{eq:orth_cond} correspond to bonds that point toward the first layer. Consequently, each node besides the root node has one upward-pointing bond only.

Plugging \autoref{eq:ttns_rootexpansion} into Schrödinger's time-independent equation (TISE),\cite{Quantisierung1926schrodinger}
using the orthogonality relations, and keeping all but the root tensor $A^{[1]}_{ij}$ fixed,  its values %
are improved by diagonalizing $\hat H$ in the basis of the configurations, i.e., diagonalizing the matrix with entries $\matrixe{\Phi_{ij}}{\hat H}{\Phi_{\tilde i \tilde j}}$.
This brings $\ket{\Psi}$ closer to the ground state.
Diagrammatically, the eigenvalue problem is depicted in \autoref{fig:tensor_overview}(c).

How do we update all the other tensors to obtain the ground state? %
It would be convenient to solve an eigenvalue problem for each tensor in the same fashion we did for the root tensor.
Indeed, this is possible  by changing the orthogonality relations and define configurations for a different tensor in the TTNS, similar to the expansion in \autoref{eq:ttns_rootexpansion}. 
This process is called canonicalization. 
To understand this, we 
view $\matr A^{[1]}$ as a non-orthogonal matrix, which we 
orthogonalize, e.g., using a QR decomposition, $\matr A^{[1]} = \matr Q \matr R \equiv \matr{\tilde A}^{[1]} \matr R$, and absorb $\matr R$ into $A^{[2,2]}_{klj}$, 
\begin{equation}
   A^{[1]}_{ij}    A^{[2,2]}_{klj} = \sum_{x} \tilde A^{[1]}_{ij}  %
   R_{jx} %
   A^{[2,2]}_{klx}  = \tilde A^{[1]}_{ij}  %
   \tilde A^{[2,2]}_{klj}.
\end{equation}
This is shown in \autoref{fig:tensor_overview}(d), next to the initial and final orthogonality relations of  $\matr A^{[1]}$ in \autoref{fig:tensor_overview}(e).
The QR decomposition changes the
orthogonalization conditions shown in \autoref{eq:orth_cond} and, consequently, the root node in the tree, which is now $\matr A^{[2,2]}$ (using the notation of numbering tensors for the original tree),
c.f.~\autoref{fig:tensor_overview}(d).

After performing the QR decomposition, we can set up an equation similar to \autoref{eq:configs}, $\ket{\Psi} = \sum_{klj} \tilde A^{[2,2]}_{klj} \ket{\widetilde \Phi_{klj}}$, and diagonalize $\hat H$ in the basis of the now orthogonal configurations $\ket{\widetilde \Phi_{klj}}$. This gives us an update to $ \tilde A^{[2,2]}_{klj}$ and a better approximation to the ground state.
We then repeat this process for each tensor in the TTNS, as shown by purple arrows in \autoref{fig:tensor_overview}(c). 
Doing this for all tensors is called a sweep in the DMRG language and is the hallmark of the DMRG algorithm.
We obtain a converged ground state for a given bond dimension by repeating the sweep multiple times.
This procedure is similar to the self-consistent field algorithm of the Hartree-Fock method, where 
molecular orbital coefficients are generated and a Fock matrix is diagonalized successively until convergence.\cite{Density2008chan}

\paragraph*{Obtaining the Hamiltonian}
As with many other methods, efficient simulations of tensor network methods require that the system-specific Hamiltonian has a suitable mathematical representation. 
Numerous simulations are based on normal coordinates that neglect kinetic coupling terms 
and use Taylor expansions as the potential energy form, e.g., for vibronic coupling models and force fields.\cite{Multimode1984koppel,Picking2024fortenberry}
This leads to a Hamiltonian that is a sum of products (SoP) of one-dimensional terms.
It can be converted to the equivalent of an MPS for an operator, dubbed matrix product operator (MPO), where the MPO tensors are diagonal along the virtual (non-physical) indices, as shown in \autoref{fig:tensor_overview}(f).\cite{Tensor2024larsson}
An operator counterpart of a TTNS is also possible, which is dubbed tree tensor network operator (TTNO) or multilayer potfit.\cite{Multilayer2014otto,Optimal2024li,Optimal2025cakir} 
These forms of the Hamiltonian can be easily integrated into TTNS machinery and lead to a low computational scaling with respect to the bond dimension. 

Given the TNS-friendly forms of Hamiltonians for model systems and force fields, we would like to have similar Hamiltonians even for non-model systems. 
Alternatives that do not require particular functional forms are possible when using quadrature,  but they are much more difficult to develop and implement.\cite{Timedependent1996manthe,Multilayer2008manthe,New2018wodraszka,Nonhierarchical2024ellerbrock}
Fortunately, 
the MCTDH community and others have shown that SoP and related Hamiltonian forms are possible also for non-model systems.
Indeed, 
the kinetic-energy operator has an SoP form for polyspherical coordinates,\cite{Exact2009gatti,Automatic2012ndong} 
or can be approximated using SoP forms for some other curvilinear coordinates.\cite{Quantum2007evenhuis}
However, in most cases, realistic potential energy surfaces (PESs) are rarely in TNS-friendly forms,
and we must (re)-fit the PES into such a form.
This is also required if we use coordinate systems that are different from the ones used to fit the PES.
In practice, there are many methods that perform such refitting.\cite{Product1996jackle,Using2006manzhos,Multilayer2014otto,Communication2014koch,Fast2015rakhuba,Transforming2020schroder,Adaptive2022aerts,Lowrank2020panades-barrueta,Neural2025hino,Systematically2026aerts} 
A particularly impressive method is the 
Monte-Carlo-based ``candecomp'' fitting procedure, which 
leads to a compact SoP form of the Hamiltonian, and which 
recently 
enabled accurate simulations for fluxional molecules such as the 33D Eigen ion.\cite{Transforming2020schroder,Stateresolved2022larsson,Coupling2022schroder} 
In our simulations,
we have used these coordinates and PES refits, and integrated them into our DMRG TTNS method.
For example, we relied on MCTDH machinery and used
polyspherical coordinates and candecomp SoP potentials for the Zundel ion and for the Eigen ion (the PESs were based on accurate fits from coupled cluster energy data\cite{Initio2005huang,Communication2017yu,Automated2019schran,Infrared2022beckmann}).
In addition, we used Radau-based
hyperspherical coordinates and an MPS fit of a nonadiabatic neural-network-based PES\cite{Diabatic2019williams} for the \ce{NO3} radical.

\begin{figure}
     \includegraphics[width=\columnwidth]{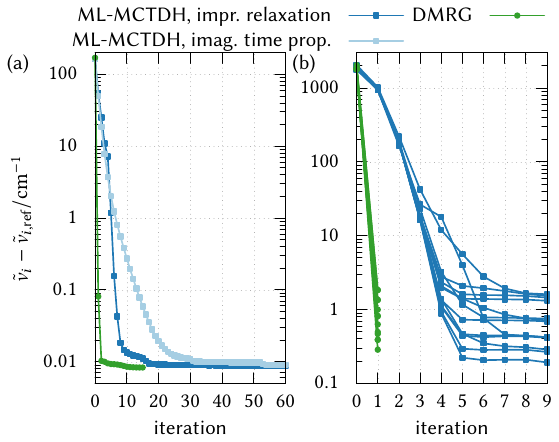}
     \caption{%
     Comparison of the DMRG algorithm with two ML-MCTDH approaches for \ce{CH3CN}: imaginary time propagation (ITP) and improved relaxation, which diagonalizes an effective Hamiltonian for the root tensor and uses ITP for all other tensors.
     (a) Absolute energy errors as a function of iteration 
     for the ground state. 
     (b) Same as (a) but for the lowest 13 states using state averaging. 
     All three algorithms start from the same random initial TTNS.
     The state averaged optimizations use the same bond dimensions as the ground state optimizations, thus the larger error.
     The iteration is proportional to the runtime and is  one sweep for the DMRG, one propagation with a  time step of $\unit[1]{fs}$ for ITP, and one diagonalization followed by ITP propagation for improved relaxation, respectively.
     Adapted  with permission from \lit{Computing2019larsson}. Copyright 2019, AIP Publishing.
     }%
     \label{fig:CH3CN_ML_vs_TTNS}
\end{figure}

\paragraph*{Relationship to the ML-MCTDH method}
Instead of using the DMRG algorithm to compute the ground state and excited states as outlined in the next paragraph,
we can also derive the complete gradient of the energy with respect to  
all TTNS tensor entries and use this to minimize the energy. 
This corresponds to solving the ML-MCTDH equations of motions for imaginary time propagation or the corresponding ``improved relaxation'' method,\cite{Iterative2014wang}
which decouples the equations  of motions of the root node with that of the other nodes.
However, the resulting equations 
contain redundancies that are very nonlinear and difficult to solve.
This is showcased in \autoref{fig:CH3CN_ML_vs_TTNS}
for computing the ground state and the first 13 eigenstates from a state-averaged computation for the 12-dimensional \ce{CH3CN} molecule.
For this example,
the DMRG TTNS algorithm requires only one (three) iterations to get to a converged ground (state-averaged) state, %
whereas the
best ML-MCTDH method requires eight (17) iterations to reach the same convergence.

\paragraph*{Optimizing excited states}
Using the DMRG or similar algorithms, we can obtain excited states in multiple ways. 
These include state averaging,\cite{Iterative2012hammer,Iterative2014wang}
as showcased in \autoref{fig:CH3CN_ML_vs_TTNS},
diagonalizing not $\hat H$ but variants such as $(\hat H-E)^2$ where the new ground state is an eigenstate of $\hat H$ that is close to a target energy $E$,\cite{Optimization2019baiardi}
as well as projecting out or shifting previously computed states in energy.\cite{Iterative1973shavitt}
We have compared some of these approaches in \lit{Computing2019larsson} and through additional applications in \lit{Stateresolved2022larsson}, and found that 
computing eigenstates one-by-one and energy-shifting previously computed states works extremely well and to high accuracy, even when thousands of lowest-energy states are computed. This procedure means
that, instead of diagonalizing $\hat H$ we instead  
diagonalize  $\hat H + \sum_I (E_I + S) \ketbra{\Psi_I}{\Psi_I}$, where $\ket{\Psi_I}$ is the previously computed state with energy $E_I$, and $S$ is a large number that separates the shifted states from the  next lowest-lying state.
Importantly, each $\ketbra{\Psi_I}{\Psi_I}$ term requires a computational cost that is smaller than that of a single product term of a SoP Hamiltonian. Thus, as long as the number of computed states is 
less than the number of terms in the used SoP Hamiltonian, the computational cost associated with the state shifting  is negligible, and it can be implemented embarrassingly parallel. 
In contrast to projecting out eigenstates, shifting them in energy is numerically more robust, because it is less prone to non-orthogonality errors  and it avoids the occurrence of a null space in the Hamiltonian.\cite{Computing2019larsson}
However, high-energy eigenstates are often harder to converge and thus require more sweeps.

\if\USEACHEMSO1
\paragraph*{Recent simulations} 
\else
\section*{Recent simulations} 
\fi
We now discuss some explicit examples of our state-shifting TTNS-based DMRG procedure and how to further improve it.
Particularly, 
we were able to compute thousands of eigenstates of fluxional vibrational systems to very high accuracy. 
This started with computing more than 1000 states of the Zundel ion, %
which we will describe in more detail below.\cite{Stateresolved2022larsson}
Later, we extended this to vibronic systems 
and computed more than 2500 states of the \ce{NO3} radical, which included five electronic states and 
covered vibrational excitations up to $\unit[6000]{\icm}$.\cite{vibronic2024larsson}
We were able to assign all 180 states up to $\unit[3000]{\icm}$ and revealed discrepancies in experimental assignments. %
For comparison, only the lowest 50 states have previously been computed and partially assigned. 

Another recent application was the computation of 5000 vibrational states of the \ce{CH3CN} molecule.\cite{Benchmarking2025larsson}
Together with a
normal-mode-based
quartic force field,\cite{Calculations2005begue,Using2011avila}
this molecule  has been established as a benchmark for methods to compute vibrational states. Compared to existing applications,  our results
increased the number of computed states by up to 5 while \emph{simultaneously} boosting the accuracy of their energies by a factor of more than  $140$.

\begin{figure}
  \includegraphics{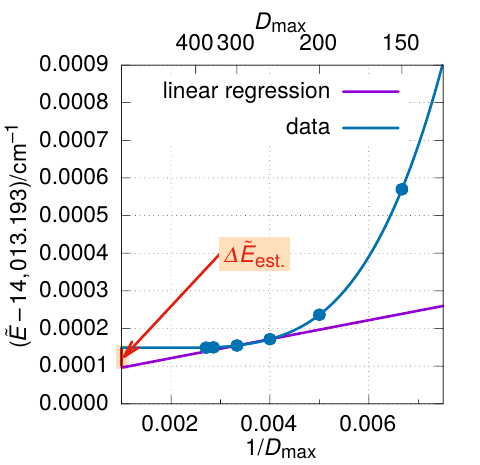}
  \caption{
  DMRG energy extrapolation and error estimate calculation.
  Shown are DMRG energies as a function of 
  the inverse max.~bond dimensions $1/\bdimmax$ (blue points),
  and a linear fit that uses  two of these energies (purple line). Its vertical intercept yields a lower bound of the energy and a corresponding error estimate.
  In practice, the fit is done for two energies with the largest bond dimensions. Here, it is shown for energies with smaller $\bdimmax$ values, giving  an error estimate that is larger than necessary.
  The blue line connects and extrapolates DMRG energies using  cubic spline interpolation.
  Reprinted from \lit{Benchmarking2025larsson}, CC BY-NC-ND 4.0 license.
  }
  \label{fig:Dextrapol}
\end{figure}

Importantly, we not only computed 5000 states but established reliable error estimates for each of them, a task that becomes increasingly more important.\cite{Uncertainty2026frombgen}
Assuming an exact Hamiltonian, thus ignoring the error of the force field and similar assumptions, 
our TTNS computations have three main error contributions. The first one is due to the DVR discretization. We estimated this by converting each state to a fully variational harmonic oscillator basis with fewer basis functions than DVR points. This error turned out to be negligible, as the TTNS wavefunction decomposition allows us to use very large basis sizes, %
since this amounts to an increase of only a single dimension in a typically two-dimensional tensor in the TTNS. 
The second error contribution is due to the finite bond dimension. Borrowing from previous work in condensed matter physics,\cite{Simulation2009tagliacozzo}
we estimated this error by extrapolating the energy, which is  a convex function of $1/\bdimmax$; see \autoref{fig:Dextrapol} for an example.
The third error contribution  is more subtle and due to small overlaps between each computed eigenstate. Among others, this non-orthogonality error introduces small energy differences of up to $\unit[0.2]{\icm}$ in states that should be degenerate.
We resolved this error by viewing
all 5000 computed states as a nonorthogonal basis that we used  in a generalized but well-conditioned eigenvalue problem. 
We showed that this procedure decreased the  
energy errors from $\unit[0.2]{\icm}$ to below $\unit[0.0007]{\icm}$ for all 5000 states.
A comparison to energies from the literature showed that even for this ``simple'' 12-dimensional quartic force field,  some previously reported energies have an error that is up to two orders of magnitude larger than anticipated. 
This highlights the challenge of computing reliable and accurate high-dimensional vibrational eigenstates.

\paragraph*{Computing infrared spectra}%

Once we have optimized the vibrational eigenstates,  we can compute IR spectra through the linear-response equation %
along one component of the dipole operator $\hat \mu$,\cite{Timedependent1990balint-kurti}
\begin{equation}
  I(E)=\frac{\pi E}{3 c \epsilon_0 \hbar} \sum_I\left|\matrixe{\Psi_I}{\hat \mu}{\Psi_{0}}\right|^2 \delta\left(E+E_0-E_I\right).
  \label{eq:ir_tise}
\end{equation}
Making use of the properties of the $\delta$ distribution, we can also obtain $I(E)$ from a Fourier transform of the autocorrelation function,  $A(t) = \matrixe{\hat \mu \Psi_0}{\exp(-\ii \hat H t /\hbar)}{\hat \mu \Psi_0} \equiv \braket{\Psi(t=0)}{\Psi(t)}$,
giving 
\begin{equation}
 I(E)=\frac{E}{3 c \epsilon_0 \hbar^2} \operatorname{Re} \int_0^{\infty} \exp[\ii \left(E+E_0\right) t / \hbar] A(t) \dd t.
  \label{eq:ir_tdse}
\end{equation}
This approach requires solving the time-dependent Schrödinger equation (TDSE) and thus is the hallmark of the ML-MCTDH method. 
In \lit{Tensor2024larsson}, we highlighted that a particular propagation algorithm of the ML-MCTDH method is identical to the TD-DMRG method, which propagates the TTNS using sweeps.
Using this algorithm with imaginary time propagation leads to the DMRG in the limit of an infinite imaginary time step.

The TISE-based approach, \autoref{eq:ir_tise}, provides eigenstate-resolved stick-spectra but requires computing many eigenstates, whereas the TDSE-based approach, \autoref{eq:ir_tdse}, provides the spectrum in a large frequency region but with a resolution limited by the total propagation time. Full wavefunctions cannot be easily extracted from the time propagation, and  it is not possible to improve the accuracy for a single peak only.
In our work, we use both TTNS-based DMRG and TD-DMRG/ML-MCTDH approaches. %
Both approaches are implemented in our own, independent code, which we have used in all of our previous work. The first TD-DMRG/ML-MCTDH application of our code was on the Zundel ion.\cite{Stateresolved2022larsson}
\paragraph*{Simulations of the Zundel ion}
Not only is the Zundel ion,  \ce{H+.(H2O)2},
key to understanding protonated bulk water and larger water clusters,\cite{Energiebander1968zundel,Nature1999marx,Quantum1997tuckerman,Coupling2022schroder,Capturing2020yang,Hydrated2016dahmsa,Largeamplitude2017dahms}
it also leads us into examining its many other intriguing features such as degenerate minima and rearrangement paths,\cite{Rearrangements1999wales}
its extreme hydrogen bonding,\cite{Crossover2021dereka} its strong sensitivity to electric fields,\cite{Extremely1972janoschek,Largeamplitude2017dahms}
as well as its strong isotope effects\cite{Isotopic2008mccunn,GasPhase2003asmis,Strong2009vendrell,Unraveling2011guasco}. %
The IR spectrum of the Zundel ion displays a  dominant doublet  around $\tilde \nu = \unit[1000]{\icm}$, which has  puzzled 
scientists for now almost  two decades.\cite{Vibrational2005hammer,Vibrational2006kaledin,Dynamics2007vendrell,Full2007vendrell,How2014rossi,Reduced2019bertaina,Stateresolved2022larsson,Revisiting2025ma}
While initial VCI and Monte Carlo simulations in 2005
on an accurate coupled-cluster-based polynomial PES\cite{Initio2005huang}
could not reproduce the doublet,\cite{Vibrational2005hammer}
in 2007,
full-dimensional, time-dependent MCTDH simulations using  polyspherical coordinates
revealed that it stems from a Fermi resonance  between a state with two quanta in the wagging motion and one quantum in the \ce{O}-\ce{O}-stretch motion, and another state with one quantum in the proton transfer motion.\cite{Dynamics2007vendrell}

15 years later, using more accurate eigenstate-resolved TTNS DMRG simulations, we showed that 
these MCTDH simulations were not fully converged and missed an important third peak when using the same PES,  resulting in a surprising triplet in the IR spectrum. In this triplet, next to the proton transfer and wagging/O-O-stretch states, a third state with four quanta in the wagging motions contributes.\cite{Stateresolved2022larsson}
However,
we also showed that
this triplet is due to subtle errors in the PES and another, neural-network-based PES developed 14 years later than the previously used PES,\cite{Automated2019schran,Infrared2022beckmann}
recovered again the doublet with only two peaks and overall a closer match to the experimental spectrum. This study highlights that small energetic changes in the PES, in this case on the order of only $\unit[40]{\icm}$, can lead to dramatic changes in the vibrational dynamics of fluxional  molecules.
In addition, this study shows how difficult it is to reliably converge the IR spectrum for such a system, both in terms of the PES generation and of the wavefunction computation.
Along a similar vein,
in 2025, new VCI simulations that included up to 13 of the 15 Zundel ion modes showed that the original VCI simulations from 2005 missed the doublet due to a basis size that was too small.\cite{Revisiting2025ma} %
However, these  reduced-dimensional simulations could not recover the triplet we found with converged, full-dimensional simulations on the same PES. %

\begin{figure}
    \includegraphics[width=\columnwidth]{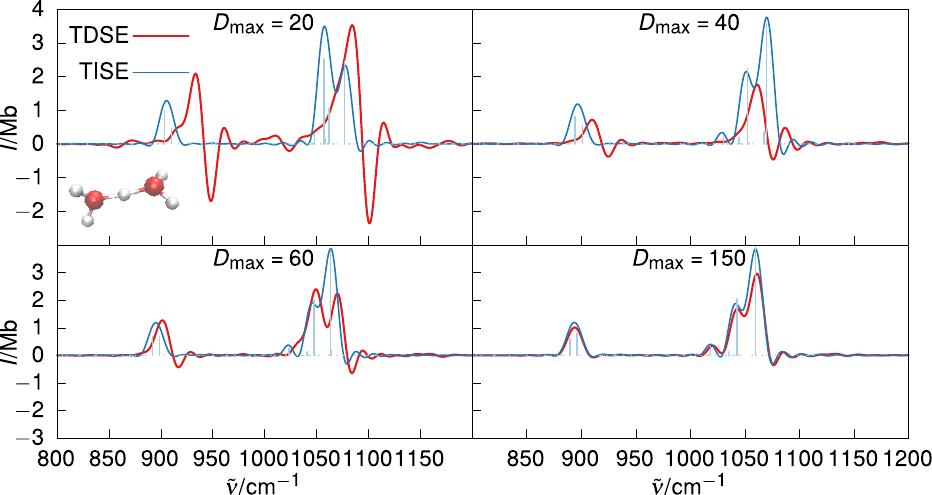}
    \caption{Zundel ion IR spectrum computed using two TTNS approaches: time-propagation (TDSE, red curves) and eigenstate optimization (TISE; blue curves and sticks) for different basis sizes/bond dimensions. 
    The total propagation time is $T=\unit[2000]{fs}$.
    Both curves use the same $\cos[\pi t/(2T)]$ kernel for convolution, leading to a scaling factor of $4T/\pi$ for the stick spectrum.}
    \label{fig:zundel_ir}
\end{figure}

In the following, we revisit the Zundel ion in the region of the doublet using the PES from 2005\cite{Initio2005huang} and our setup from \lit{Stateresolved2022larsson}
to highlight the difficult convergence of the IR intensities.  (We omit here errors in the intensities due to the separation of the rotational from the vibrational motions.\cite{Use1992lesueur})
We use both time-independent and time-dependent approaches to compute the spectrum. 
The time-independent results use the DMRG algorithm, followed by a diagonalization of the generated TTNS basis.
\autoref{fig:zundel_ir} displays the comparison of the TDSE-based and TISE-based IR spectrum around $\unit[1000]{\icm}$ for different bond dimensions.
Astonishingly, the triplet is not visible in the TDSE-based IR spectrum when using $\bdimmax=20$. It only clearly appears for $\bdimmax=60$, but with qualitatively  wrong intensities,
compared to the converged $\bdimmax=150$ reference.
Consequently, the MCTDH simulations from 2007 missed the triplet due to a too small basis size.
Note that this is by no means a criticism of this heroic breakthrough MCTDH simulation,
which, back then, was one of the largest and most difficult MCTDH application that had ever been performed. %
The subtle basis convergence effects we discuss here could not have been recognized in 2007 with the available computational resources and algorithms.\footnote{The simulations from 2007 used a different approximation of the original PES, but we found that the same issue appears regardless of the used approximation.}
The basis sizes/bond dimensions we use are much larger than what is typically anticipated in ML-MCTDH/TTNS applications, but they are required for fully converging the IR spectrum and for revealing all subtle, but important features.  While the triplet is a peculiar feature of the used PES,  
quasi-exact quantum dynamics methods should still be able to reliably produce such features. 

We note that a time-independent Hamiltonian should produce a purely positive-valued IR spectrum, however large negative intensities are seen in the TDSE-based spectrum when using a small $\bdimmax=20$. The observed negative intensities are due to the time-dependent TTNS tensors, which results in a time-dependent basis (the configurations in \autoref{eq:configs}) and an effective, time-dependent Hamiltonian. The larger $\bdimmax$ is, the more accurate the basis, the smaller the time-dependency of the Hamiltonian, and the smaller the negative IR intensities. Remaining small negative contributions are due to a convolution with a $\cos$ function.\cite{Extracting1998beck}

In contrast to the TDSE-based simulations, the TISE-based simulations are more accurate, and a triplet already appears for $\bdimmax=20$.
The IR intensities are almost converged  for $\bdimmax=40$. 
This faster convergence 
is due to each eigenstate being described by an individual TTNS, whereas for the TDSE the spectral information is extracted from a time propagation of a single individual TTNS, which requires large bond dimensions for long propagation times.

\begin{figure}
\includegraphics[width=\columnwidth]{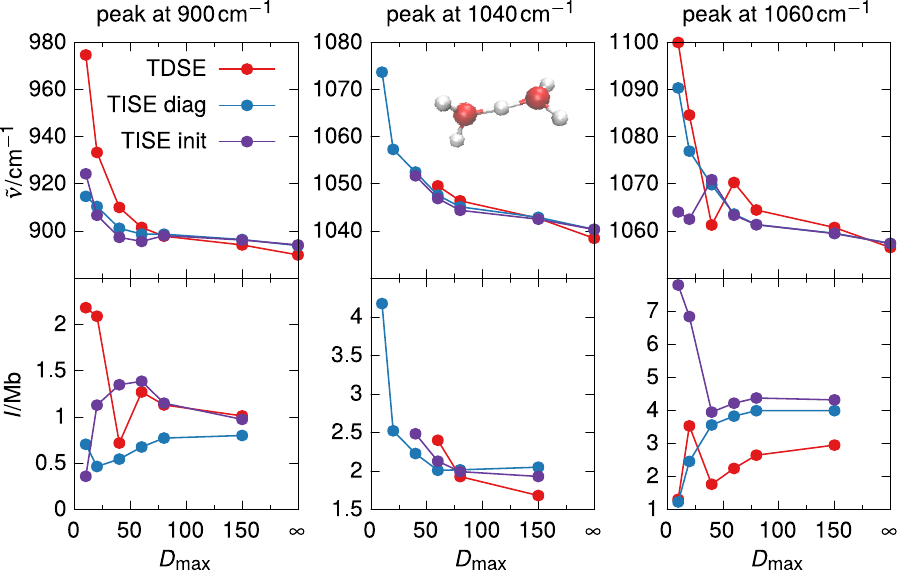}
    \caption{Convergence analysis of the triplet in the Zundel IR spectrum that appears when using the PES from \lit{Initio2005huang}.
    First row: Excitation energies taken from the peaks of the time-propagation-based spectrum in \autoref{fig:zundel_ir} (red  points), 
    from the eigenstate computation after (blue  points) and before (purple points) diagonalization.
    The last point is based on extrapolation, compare with \autoref{fig:Dextrapol}.
    Second row: Same as first row but for the IR intensities.
    Missing data for the peak at $\unit[1040]{\icm}$ means that the simulation did not lead to a significant peak in this region.
    }
    \label{fig:zundel_energies}
\end{figure}

In \autoref{fig:zundel_energies} we plot the energies and IR intensities as functions of $\bdimmax$. 
This plot also includes eigenstate-based results without additional diagonalization. The corresponding TTNSs have small overlaps to other TTNSs, which worsens the accuracy. Importantly, without diagonalization, the peak at $\unit[1040]{\icm}$ only appears for $\bdimmax\ge40$, whereas after diagonalization of the same basis, it is visible already at $\bdimmax=10$.
For larger bond dimensions, starting with $80$, the TTNSs approach the exact eigenstates and the effect of the diagonalization becomes negligible.
 
\begin{figure}
  \includegraphics[width=\columnwidth]{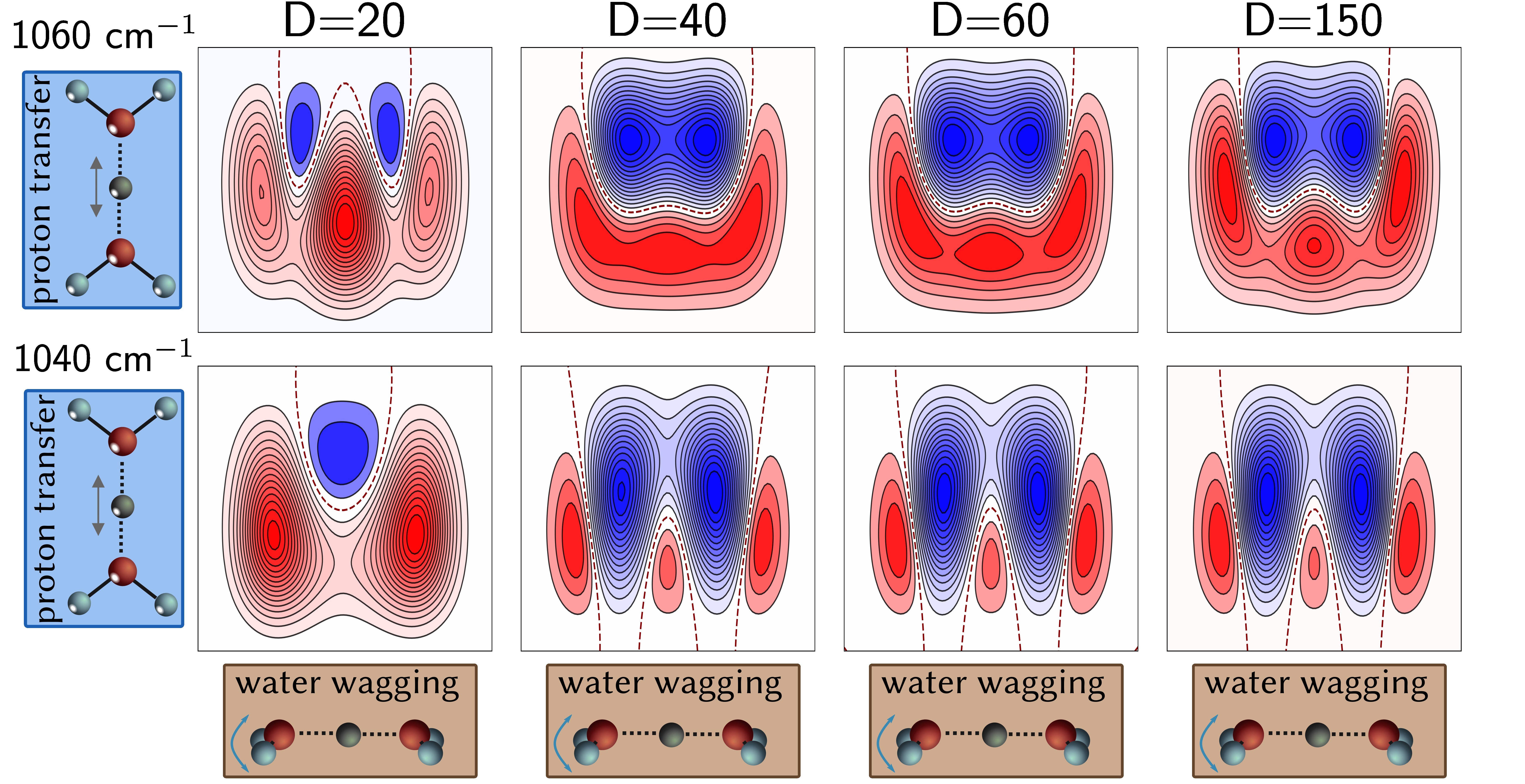} %
  \caption{Zundel ion eigenstate cuts along the proton transfer and one of the water wagging motions for different bond dimensions. Two eigenstates with energies at $\unit[\sim1060]{\icm}$ (upper row) and $\unit[\sim1040]{\icm}$ (lower row) are shown, respectively.
    The red (blue) regions correspond to the positive (negative) values. 
    The red dashed lines denote the zero contours.
  }
  \label{fig:zundel_wfcuts}
\end{figure}

In general, we get energies that converge smoothly and rapidly with bond dimension. In contrast, the convergence of the IR intensities are more difficult. 
Variational methods can give accurate energies for poor wavefunctions, and the energy error depends only to the second order on the wavefunction error.\cite{Molecular2013helgaker}
This is also demonstrated in \autoref{fig:zundel_wfcuts} where we plot wavefunction cuts for different bond dimensions after diagonalization. 
The converged $\bdimmax=150$ cuts highlight the resonance structure with one quantum (zero-crossing) along the proton transfer motion and  zero, two or four quanta along the wagging region, depending on the value of the proton transfer coordinate.
For $\bdimmax=20$, however,
even though the triplet feature is visible in the IR spectrum,
the corresponding wavefunction is qualitatively incorrect.
Qualitative convergence  is achieved for $\bdimmax=40$, 
but finer details of the wavefunction  around $\unit[1060]{\icm}$ are only converged for $\bdimmax=150$. 
This highlights the difficulty of converging observables other than energies for such complex, fluxional molecular systems.

\paragraph*{Simulations of the Eigen ion}
In addition to the Zundel ion, the Eigen ion has been regarded as one of the limiting structures of bulk protonated water.\cite{Uber1954wicke,Resolving2021calio,Spectral2005headrick,Demystifying2021zeng,Nature1999marx,Unique2011stoyanov,Ultrafast2015thamer} %
As for the Zundel ion, simulating and quantitatively reproducing the experimental spectrum is difficult.
VCI simulations in reduced dimensionality  that neglected couplings between modes 
resolved an important question about which isomer the experiment actually had measured.\cite{Communication2017yu,HighLevel2017yu,Deconstructing2018esser}
However, these simulations could not fully reproduce the experimental spectrum, in particular its low-energy region and 
a wide broadening of a peak attributed to the excess proton around $\unit[2600]{cm^{-1}}$. 
Recent breakthrough ML-MCTDH simulations of the fully coupled 33-dimensional system
using polyspherical coordinates and a candecomp PES re-fit
could reproduce the experimental spectrum for most peaks.\cite{Coupling2022schroder}
These simulations were partially limited by convergence issues, as this 33-dimensional wavefunction requires large bond dimensions.
Recently, we used the setup from \lit{Coupling2022schroder} to compute the first 1300 eigenstates with bond dimensions as large as $150$.\cite{Computing2025rano}

One important aspect of such simulations are optimized tree structures for the TTNS. As we show below, this not only dramatically improves  the efficiency of the method and the convergence issues encountered in \lit{Coupling2022schroder}, but also provides physical insights about modal couplings. %
In many cases, finding optimal tree structures are guided by 
chemical intuition and trial and error.\cite{Multilayer2011vendrell,Reaction2012welsch,Comparison2024dorfner,Further2025li} %
To make this %
less cumbersome,
we introduced a systematic and fully automated way to find optimal trees, which uses an initial tree and optimizes its structure, e.g., by permuting tensors and by adding additional tensors to the tree.\cite{Computing2019larsson,Tensor2024larsson} 
Approaches related to our optimization are now being used successfully in condensed matter physics.\cite{Automatic2023hikihara}%
Another approach is to 
find optimal trees manually by analyzing  correlations from classical molecular dynamics simulations,\cite{Optimal2023mendive-tapia}
which avoids the need of an initial guess.

\begin{figure}
\centering
\includegraphics[]{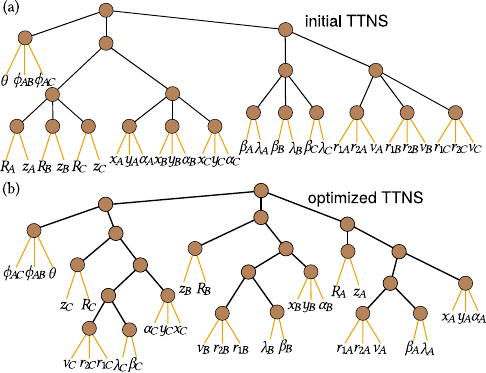}
\caption{Different tree structures for the 33-dimensional vibrational Eigen ion.
(a) Tree structure used in \lit{Coupling2022schroder}%
(b) Optimized structure from \lit{Tensor2024larsson} using the algorithm from \lit{Computing2019larsson}. 
The symbols denote the specific coordinates.\cite{Coupling2022schroder}
The indices $A$, $B$, and $C$ denote the three water units of the Eigen ion.
The mode combinations (groups of coordinates) have not been optimized during the tree structure optimization, but this is possible.
}
  \label{fig:eigen_ttns_opt}
\end{figure}

In \autoref{fig:eigen_ttns_opt} we compare the Eigen ion 
tree used in \lit{Coupling2022schroder}
with our  optimized tree from \lit{Tensor2024larsson}.
Two aspects of this comparison, which were not clear when the study in \lit{Coupling2022schroder} was performed,
are particularly important.
The first is that our optimized tree is based on three-dimensional tensors only, which reduces the scaling with respect to the bond dimension and thus the computational effort. 
Such trees are also favored in condensed matter physics and electronic structure theory.\cite{T3NS2018gunst}
Note that we used the tree with four-dimensional tensors from \lit{Coupling2022schroder} as a starting point of our tree optimization. The algorithm automatically modified the initial tree to yield
more optimal three-dimensional tensors. %
The second aspect about the comparison is that the optimized tree clusters coordinates that belong to the same water subunit in the Eigen ion.
This reveals that the coordinates within each water subunit are strongly coupled, which for such a complex, fluxional protonated water cluster is a nontrivial physical insight.

\begin{figure}
  \includegraphics[scale=.8]{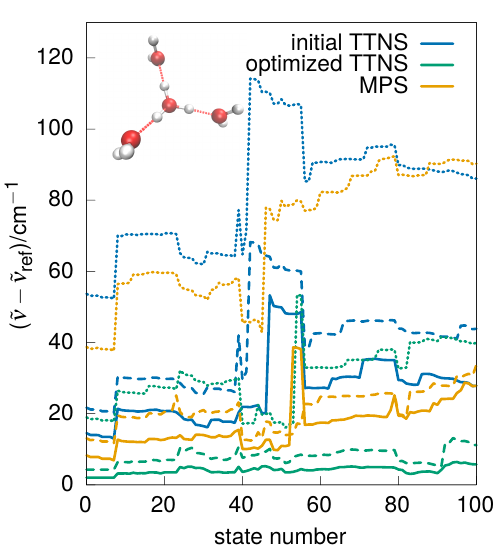}
  \caption{Error of the first 100 energies  of the 33-dimensional Eigen ion for the TTNS tree structures from \autoref{fig:eigen_ttns_opt} and for an MPS with the same coordinate ordering as the optimized tree.
  Straight, dashed, and dotted lines correspond to a used bond dimension of $70$, $50$, and $20$, respectively.
  The references energies use a bond dimension of $150$.}
  \label{fig:eigen_Dcomparison}
\end{figure}

While the tree in \autoref{fig:eigen_ttns_opt} was optimized for a ground state wavefunction, is the tree also good for excited states? 
We address this question 
in \autoref{fig:eigen_Dcomparison}
by comparing the energy errors of the first 100 states for different bond dimensions with respect to the trees  from \autoref{fig:eigen_ttns_opt} and to an MPS where the coordinates are ordered based on our optimized tree.
Strikingly, for all 100 eigenstates, our optimized tree with $\bdimmax=50$ results in a smaller energy error than both the initial, unoptimized tree and the MPS, even when these  use a larger $\bdimmax=70$. 
Furthermore, the MPS with optimal coordinate ordering performs much better than the initial tree and for $\bdimmax=70$ leads to a max.~energy error that is $\unit[15]{\icm}$ smaller than that of the initial tree.
Consequently, MPSs can outperform TTNSs if the tree structure is not optimized. 
With our optimized tree and $\bdimmax=70$, we obtain a max.~energy error of $\unit[6]{\icm}$,
compared to our $\bdimmax=150$ reference. %
As discussed above, the error can be improved by energy extrapolation and additional diagonalization.

\if\USEACHEMSO1
\paragraph*{Beyond the Density Matrix Renormalization Group}
\else
\section*{Beyond the Density Matrix Renormalization Group}
\fi

Our state-shifting DMRG-based TTNS method 
has the advantage that the whole eigenspectrum 
up to a given energy is computed.
However, this is inefficient for simulating IR spectra, as many states are not IR active: There are 380 Zundel states up to $\tilde \nu = \unit[1500]{cm^{-1}}$ but only $\sim$20 of them are IR active!
Another issue with the state-shifting approach is a curse of dimensionality with respect to the number of states. The larger the molecule, the more low-energy motions and the larger the density of states even at relatively small energies. For the 
33-dimensional Eigen ion, there are already more than 1300 states up to 
to an excitation energy of $\unit[\sim350]{\icm}$.\cite{Computing2025rano} 
In practice, our state-shifting approach reaches its limit for 2000 to 5000 states. %

To overcome this deficiency, it is better to  directly optimize eigenstates at a given target energy. %
While the DMRG can be modified to select high-energy states,\cite{Targeted2007dorando}
we found that such approaches are stable only for the first few excited states in energy regions with very low  density of states.
A better approach is to find excited states with methods other than the DMRG algorithm. As we will show below, some appealing aspects of the DMRG can still carry over to other methods.

Well-established for computing interior eigenstates close to a target energy $\sigma$ are methods based on the shift-and-invert transformation, $(\hat H - \sigma \hat 1)^{-1}$, which turns the interior eigenstates into exterior ones, making them easier to compute. The shift-and-invert transformation has been used to compute vibrational states using the FEAST method and MPSs in \lit{ExcitedState2021baiardi} and using a block iteration method and different tensor decompositions in   \lit{Computing2021kallullathil}. %
In \lit{Computing2025rano} we introduced a different shift-and-invert method that uses the inexact Lanczos algorithm,\cite{New2000huang}
which creates a basis of vectors by \emph{approximately} (or inexactly) applying  $(\hat H - \sigma \hat 1)^{-1}$ onto an initial guess multiple times. This basis, consisting of only a dozen of vectors, is then used to diagonalize the Hamiltonian. 
In our case, we do not have vectors but TTNSs. This requires us to recast vector algebra operations such as additions and operator multiplications into approximate TTNS operations (this is also required for the aforementioned alternative methods). %
Luckily, the sweep algorithm  of the DMRG that solves the eigenvalue problem by keeping all but one tensor fixed can also be used for other operations, including TTNS addition and solving the linear system $(\hat H - \sigma \hat 1) \ket{\Psi} = \ket{\Phi}$.
Using these sweep methods and multiple adaptions to the   original inexact Lanczos procedure that take into account that every vector operation is approximate, we were able to apply the resulting TTNS inexact Lanczos method to three challenging computations: (1) 122 states in two different energy regions of \ce{CH3CN}, (2) the Fermi resonance states of Zundel ion, and (3) excited states of the Eigen ion.
We will now review the last two.

\begin{figure}[!htbp]
  \includegraphics[width=\columnwidth]{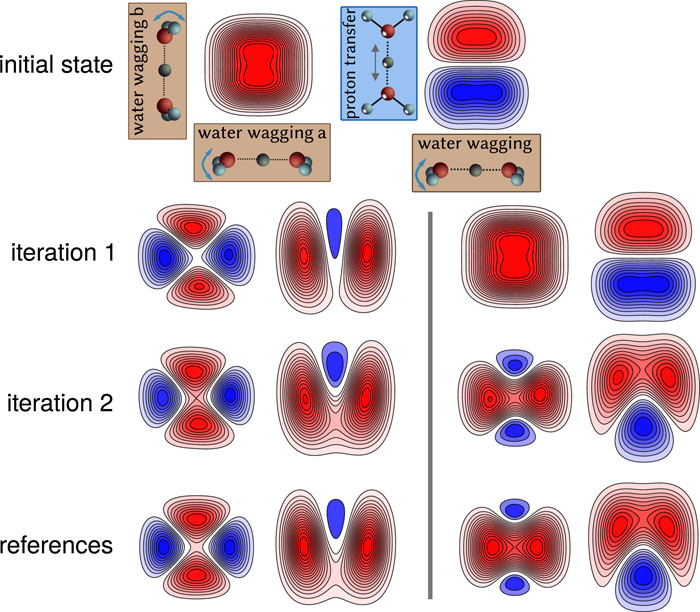}
    \caption{Inexact Lanczos optimization for the Fermi resonance in the Zundel ion. %
    The first row shows the wavefunction cuts of the initial guess state  along the two water wagging coordinates, and along the proton transfer and one of the water wagging coordinates. 
    The second and third rows display cuts of the resulting eigenstates in iterations 1 and 2 (the third state in iteration 2 is not shown). %
    We use $\bdimmax=50$ and $\sigma=\unit[1000]{cm^{-1}}$. 
    The last row shows DMRG references with the same $\bdimmax$ value. %
    Note that the used PES\cite{Automated2019schran} is different from that used in \autoref{fig:zundel_wfcuts}.
     Reprinted  with permission from \lit{Computing2025rano}. Copyright 2025, AIP Publishing.
    }
  \label{fig:zundel}
\end{figure}

While the doublet states in the Zundel IR spectrum can quickly be associated with the proton transfer excitation, the Fermi resonance contribution by a wagging/O-O-stretch state is nontrivial. %
In \lit{Computing2025rano} we showed that the TTNS inexact Lanczos method is able to target these eigenstates directly with a poor initial guess that contains a proton transfer excitation but no excitations in neither the waggings nor the O-O stretches. 
\autoref{fig:zundel} shows the 
initial guess and the first two Lanczos iterations. 
Remarkably, the method reveals the contributions of the Fermi resonance already after the first iteration.
The second Lanczos iteration refines  the state and reproduces the second state that contributes to the Fermi resonance.
As only two applications of $(\hat H - \sigma \hat 1)^{-1}$ onto a poor initial guess suffice to reproduce the Zundel doublet states,
the procedure will be very useful in quickly identifying the nature of peaks in IR and other spectra.

\begin{figure}[!tbp]
\includegraphics[width=0.92\columnwidth]{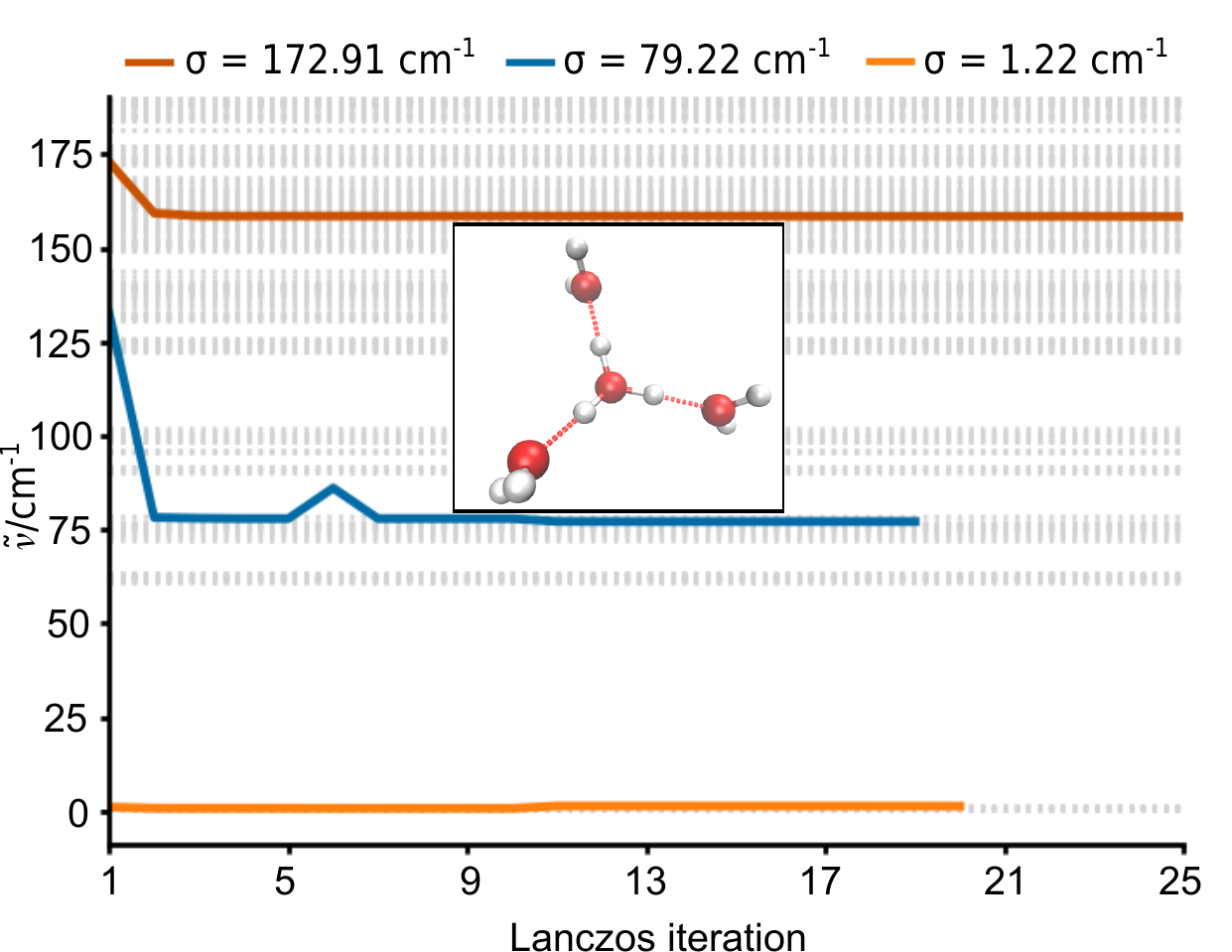}
\caption{Convergences of excited states for the 33-dimensional Eigen ion using the TTNS inexact Lanczos method and $\bdimmax=70$. 
The dotted gray lines are TTNS-DMRG  reference energies.
     Reprinted  with permission from \lit{Computing2025rano}. Copyright 2025, AIP Publishing.
}
        \label{fig:eigen_conv}
\end{figure}

In \autoref{fig:eigen_conv} we target three excited states of the full-dimensional Eigen ion.
These states were chosen based on their region in the spectrum. The targeted state with lowest energy is close to the ground state and the first excited states whereas the targeted state with highest energy is in a region with a high density of states, namely 18 states within $\unit[5]{\icm}$. 
All states exhibit strong coupling, which requires us to use a large bond dimension of $\bdimmax=70$. 
Similar to the Zundel ion optimization, after just two Lanczos iterations, the energies are converged  to the eye.
This is the case even when the density of states is very large, as shown by the ocre curve in \autoref{fig:eigen_conv}.
This opens the door to directly targeting important high-energy states such as those with \ce{O-H}-stretch excitations in large protonated water clusters and similarly complex \mbox{systems}.

\if\USEACHEMSO1
\paragraph*{Summary and Conclusions}%
\else
\section*{Summary and Conclusions}%
\fi
This Perspective outlined tree tensor network state (TTNS) methods based on the density matrix renormalization group (DMRG) for computing thousands of accurate, full-dimensional vibrational and vibronic eigenstates, with a focus on fluxional protonated water clusters as large as the 33-dimensional Eigen ion.
We showed that great care has to be taken to reliably estimate energy errors and to converge observables such as the infrared spectrum. Further, optimizing the %
tree structure used in TTNS and
multilayer
multiconfiguration time-dependent Hartree  (ML-MCTDH)
methods can lead to simulations orders of magnitude more efficient.
Lastly, we highlighted one particular avenue to go beyond traditional DMRG approaches to directly target excited eigenstates.

Our work also highlights the importance of cross-fertilization of ideas from different communities. 
The DMRG stems from condensed matter physics whereas the ML-MCTDH method stems from molecular quantum dynamics. 
Both methods share the same class of wavefunction ansatz and scientists working on both methods found various solutions to similar problems. 
More interactions and cross-fertilization will be fruitful for both communities.

Many tasks remain to be completed.
Since we can now routinely compute thousands of eigenstates, we need to find automated ways to analyze the vast amount of data in order to gain new insights into the quantum effects of molecular motion.
This might include new ideas from unsupervised machine learning and old ideas from random matrix theory.\cite{Confirmation1988zimmermann}
Computing thousands of states needs to be complemented by targeted eigenstate computations to only retrieve states with desired properties.
In analogy to post-DMRG methods in electronic structure theory,\cite{Communication2014sharma,Multireference2013saitow,PostDensity2022cheng,Matrix2022larsson,Chromium2022larsson,LargeScale2022barcza}
and following related work on vibrational active spaces,\cite{Secondorder2013mizukami,Multireference2014pfeiffer,Vibrational2023hellmers}
weakly coupled vibrations of very large molecules could be treated perturbatively.
Extending the approach to the inclusion of
rotational motion would provide an avenue for other important applications, e.g., for astrochemistry. 
While this Perspective focused on the eigenstate computation itself, a central problem is the set-up of the Hamiltonian. This includes electronic structure computations, PES fitting, PES-refitting into TTNS-friendly forms, setting up efficient coordinate systems and finding the kinetic energy operator in the chosen coordinate system in a TTNS-friendly form.
The past few years showcased great progress in these directions that will further facilitate fully quantum dynamics simulations of complex molecular systems.
\if\USEACHEMSO1
\begin{acknowledgement}
\else
\acknowledgements
\fi
This work was supported by the US
National Science Foundation (NSF) via grant no.~CHE-2312005.
This research was conducted using the Pinnacles cluster (NSF MRI, no.~2019144)
and using CENVAL-ARC compute resources on the Pinnacles cluster (NSF no.~2346744)
at the Cyberinfrastructure and Research Technologies (CIRT) at the University of California, Merced.
\if\USEACHEMSO1
\end{acknowledgement}
\fi

\end{document}